\documentclass{iau} 
\usepackage{graphicx}

\title[Statisical Challenges in the Galactic Center] 
{Statistical challenges in fitting stellar orbits around the supermassive black hole at the Galactic center}

\author[Gregory D. Martinez et. al.]   
{Gregory D. Martinez$^1$,
Kelly Kosmo$^1$,
Aurelien Hees$^1$,
Joseph Ahn$^1$,
\and Andrea Ghez$^1$}

\affiliation{$^1$UCLA Department of Physics and Astronomy \\ 430 Portola Plaza, Box 951547 Los Angeles, CA. 90095-1547\\ email: {\tt gmartine@astro.ucla.edu} \\[\affilskip]}

\pubyear{2016}
\volume{322}  
\jname{International Astronomical Union Proceedings Series}
\editors{Editor: Roland Crocker}
\begin{document}

\maketitle

\begin{abstract}
Over two decades of astrometric and radial velocity data of short period stars at the Galactic center have the potential to provide unprecedented tests of General Relativity and insight into the astrophysics of supermassive black holes.  Fundamental to this is understanding the underlying statistical issues of fitting stellar orbits.  Unintended prior effects can obscure actual physical effects from General Relativity and the underlying extended mass distribution.  At the heart of this is dealing with large parameter spaces inherent to multi-star fitting and ensuring acceptable coverage properties of the resulting confidence intervals within the Bayesian framework.   This proceeding will detail some of the UCLA Galactic Center Group's analysis and work in addressing these statistical issues.
\end{abstract}

\firstsection 
\

With over two decades of orbital astrometric and radial velocity data, the opportunity to study the properties and effects of a super-massive black hole at the Galactic Center has never been more accessible.  Unfortunately, despite having over decades worth of data, most stars at the Galactic Center do not have complete orbital phase coverage.  Within the Bayesian statistical framework, this may cause prior assumptions to bias inferred quantities and produce inaccurate confidence intervals.  Given a hypothetical large collection of datasets, the percentage of possible datasets whose derived confidence intervals contain the ``true'' physical value is defined by the quoted confidence interval.  For example, a 68\% confidence interval gives a 68\% chance that an observed dataset will produce a confidence interval that contains the true value.  Unfortunately, most algorithms used to calculate confidence intervals do not guarantee that actual calculated confidence is the same as the quoted confidence in the regime where data are not rigorously constraining.  This is especially true for datasets with incomplete orbital phase coverage where prior information can have a profound impact on the resulting confidence intervals (\cite{Lucy14}).  Thus, it is not unreasonable to expect data sets with incomplete orbital phase coverage to produce inaccurate confidence intervals.  We investigate this by calculating the statistical efficiencies, defined as the fraction of confidence intervals that cover the assumed true value as compared to the defined confidence level, as well as the statistical bias, as defined by the fraction of times the assumed true value is above the median.  In figure 1, we compare confidence intervals of the mass of, as well as the distance to, the super-massive black hole in the Galactic center derived using three different prior assumptions:  uniform priors in model orbital parameters (blue dots) and two different priors uniform in the observable (not model) parameter space (green triangles and red crosses).  We investigate confidence intervals derived in eight test cases of varying coverage as described in Figure 1.  For datasets with complete orbital phase coverage, prior assumptions have little effect over the resultant confidence intervals.  However, inferred median values seem to be consistently biased when compared to the true value in tests with partial orbital phase coverage.  As shown in Figure 1, our observable-based priors help reduce the magnitude of this effect, though they do not eliminate it in all cases.  Of the prior assumptions tested, priors that assume uniform astrometric and radial velocity observables produce the least biased confidence intervals and generally have statistical efficiencies close to or above one.  Thus, we conclude that in cases where data are not rigorously constraining, prior assumptions assuming uniformity in observables produce more accurate, and less biased, confidence intervals.

%
\begin{figure}[b]
\begin{center}
\rotatebox{0}{\includegraphics[width=0.45\hsize]{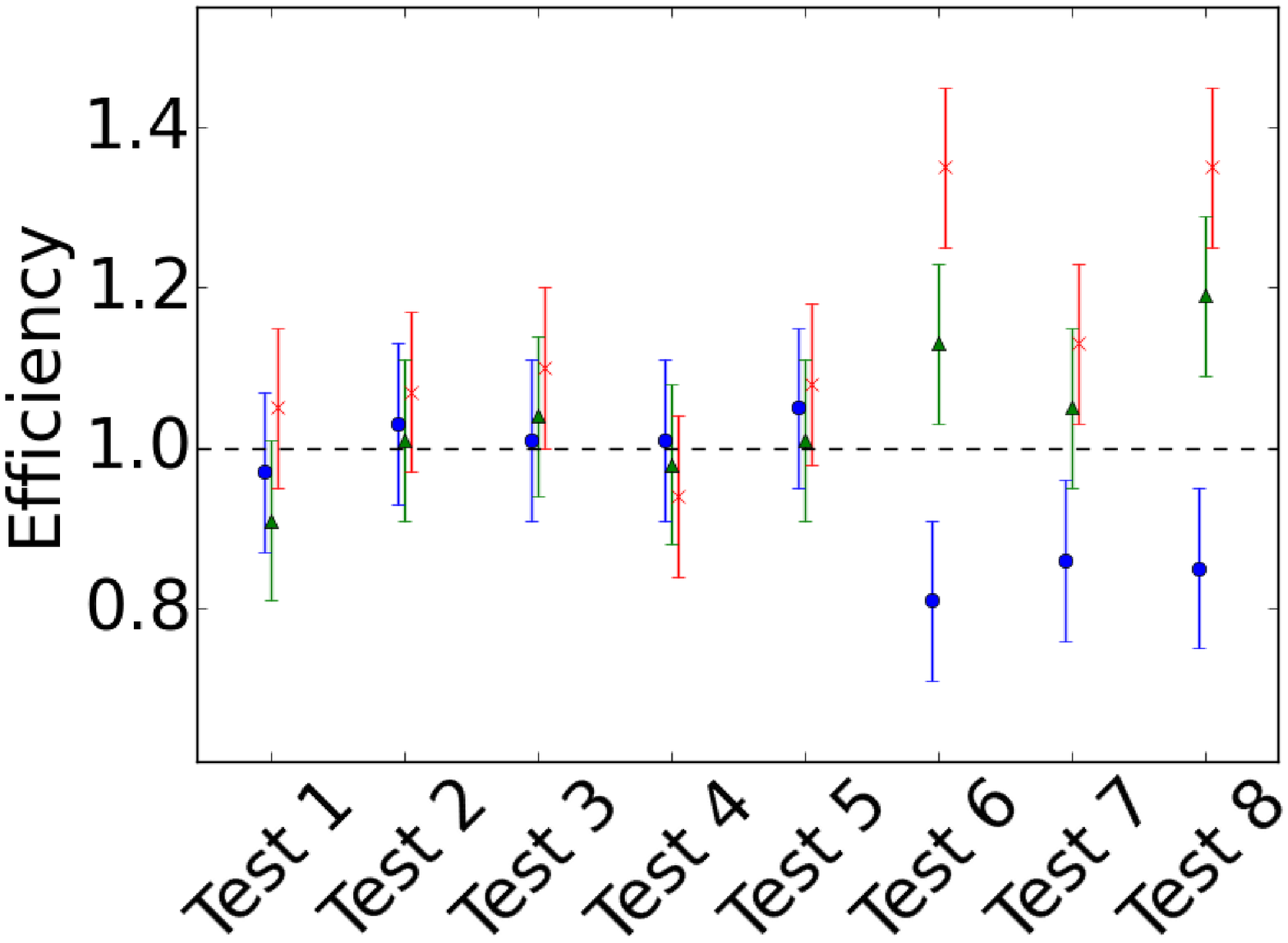}}
\rotatebox{0}{\includegraphics[width=0.45\hsize]{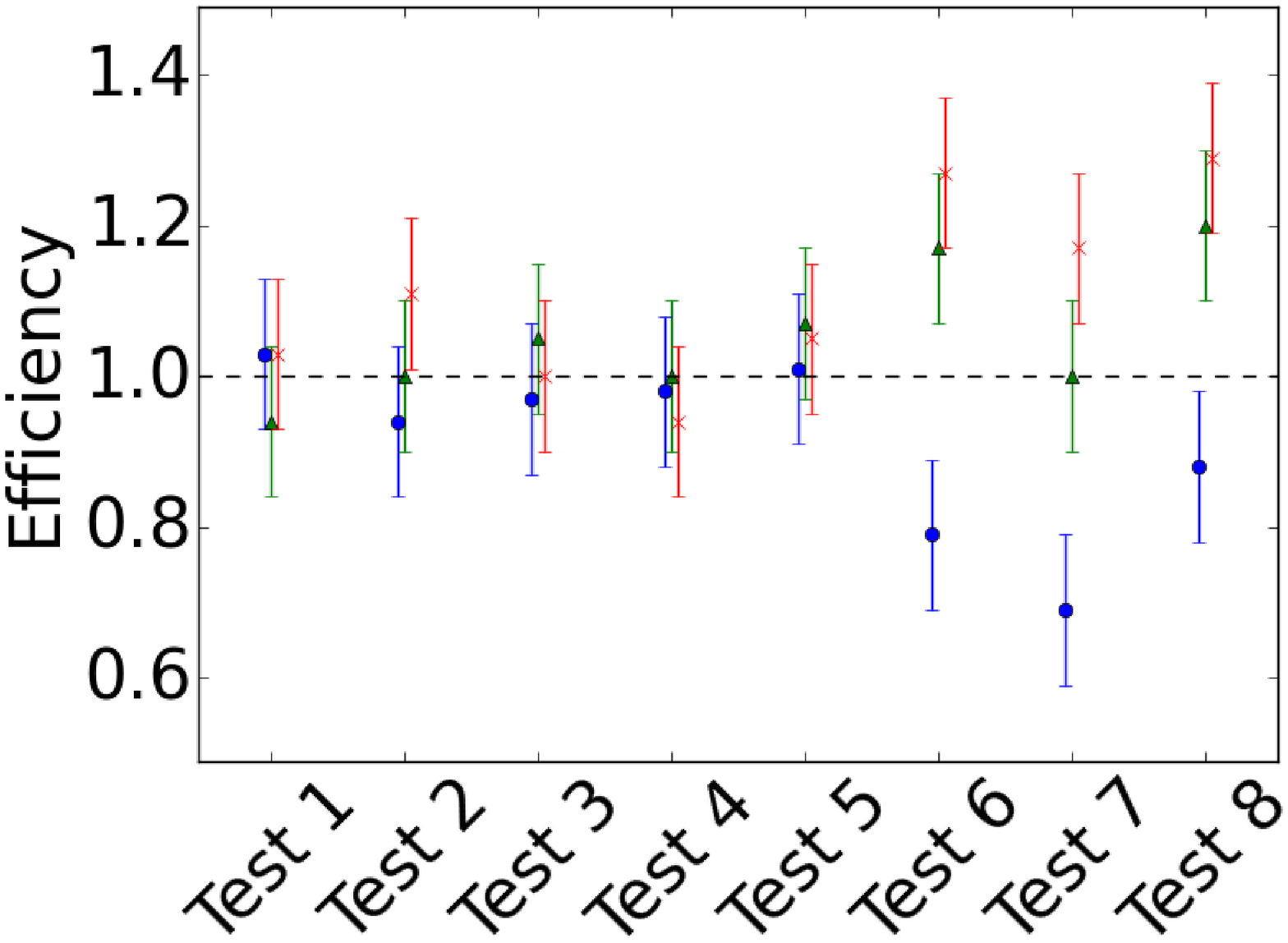}}
\rotatebox{0}{\includegraphics[width=0.45\hsize]{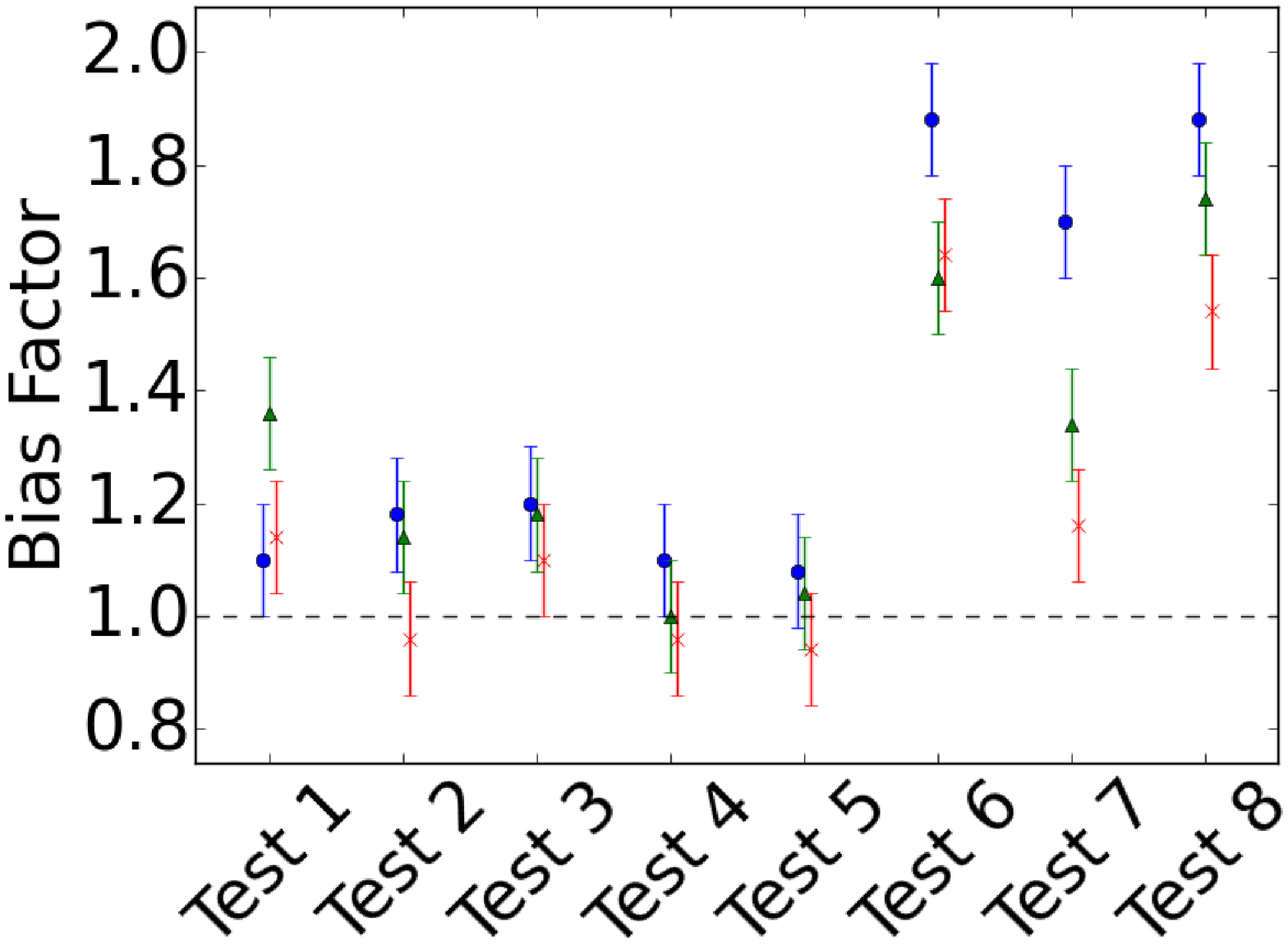}}
\rotatebox{0}{\includegraphics[width=0.45\hsize]{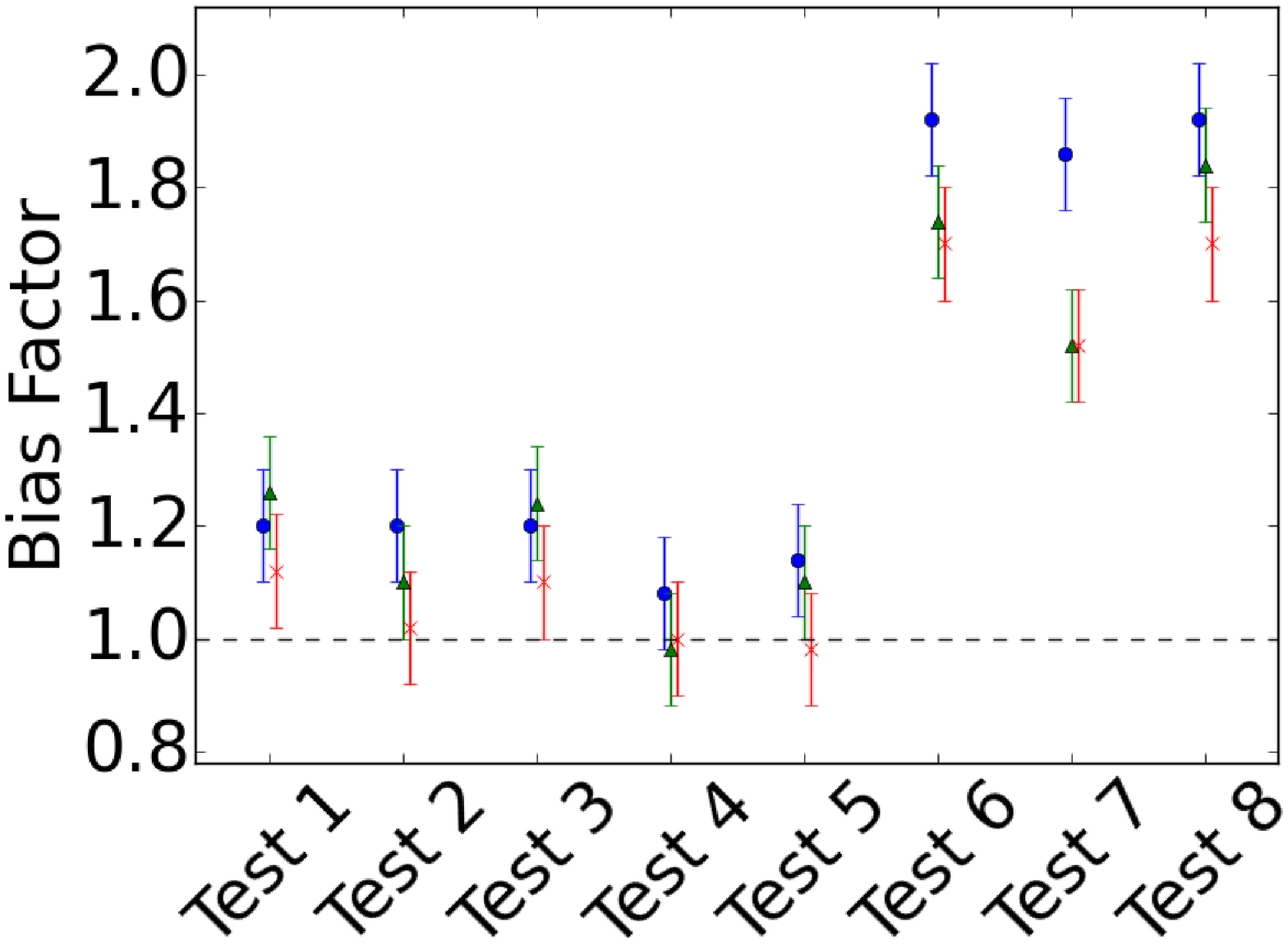}}
 \caption{
 Shown are the Statistical efficiencies (top panels) and biases (bottom panels) of various Bayesian confidence intervals for both the mass of the super-massive black hole in the Galactic center (left panels) and the distance to Galactic center (right panels). The efficiencies and biases are calculated from 100 mock data sets assuming a reasonable set of ``true'' parameters.  Plotted in the top panels are the fraction of confidence intervals that cover the assumed true value as compared to the defined confidence level.  
 Plotted in the bottom panels are the fraction of times the ``true'' value is above the median in various tests.  We compare confidence intervals derived using three priors:  uniform priors in model parameters (blue dots), uniform in astrometric observables (green triangles) and uniform in both astrometric and radial velocity observables (red crosses).  Plotted are the efficiencies for eight test cases:  datasets with the same distribution as S0-2 data (test 1, \cite{Boehle16}), with even full phase coverage (test 2), and with partial coverage near periapse or apoapse, respectively, using: actual measured errors for S0-2 (test 3 (periapse) and 6 (apoapse)), increased RV errors (test 4 (periapse) and 7 (apoapse)), and decreased frequency of RV data (test 5 (periapse) and 8 (apoapse)).}
   \label{fig1}
\end{center}
\end{figure}

\end{document}